\documentclass[preprint,proceedings]{rmaa}
\usepackage{paralist}

\usepackage{psfrag,color}

\def\erg{{\rm\thinspace erg}}

\def\K{{\rm\thinspace K}}
\def\keV{{\rm\thinspace keV}}
\def\km{{\rm\thinspace km}}
\def\kpc{{\rm\thinspace kpc}}

\def\Msun{\hbox{$\rm\thinspace M_{\odot}$}}

\def\s{{\rm\thinspace s}}
\def\yr{{\rm\thinspace yr}}

\def\ergps{\hbox{$\erg\s^{-1}\,$}}

\def\kmps{\hbox{$\km\s^{-1}\,$}}

\def\Msunpyr{\hbox{$\Msun\yr^{-1}\,$}}

\SetYear{2002}
\SetConfTitle{Galaxy Evolution: Theory and Observations}

\title{Cluster cores and cooling flows} 

\author{
  A.C. Fabian,\altaffilmark{1}} 

\altaffiltext{1}{Institute of Astronomy, Cambridge, U.K.}

\shortauthor{Fabian}
\shorttitle{Cooling Flows}

\fulladdresses{
\item A.C. Fabian, Institute of Astronomy, Madingley Road, Cambridge CB3
0HA, U.K.}

\indexauthor{Fabian, A.C.}
\abstract{
The gas temperature in the cores of many clusters of galaxies drops
inward by about a factor of three or more within the central 100~kpc
radius. The radiative cooling time drops over the same region from 5
or more Gyr down to below a few $10^8\yr$. Although it would seem that
cooling flows are taking place, XMM-Newton spectra show no evidence
for strong mass cooling rates of gas below 1--2 keV. Chandra images
show holes coincident with radio lobes and cold fronts indicating that
the core regions are complex. The observational situation is reviewed.
It is likely that some form of heating is important in reducing the
mass cooling rates by a factor of up to ten. Conduction of heat from
the hot outer gas or heating from a central radio source are
discussed, together with possible ways in which continued cooling
might occur. Parallels with the formation of the baryonic part of
galaxies are explored.}

\addkeyword{Clusters of Galaxies}
\addkeyword{Cooling flows}
\addkeyword{Galaxy formation}

\begin{document}
% Typeset article header
\maketitle

\section{Introduction}

The gas density within the central 100 kpc or so of the centre of most
clusters of galaxies is high enough that the radiative cooling time of
the gas is less than $10^{10}$~yr. The cooling time drops further at
smaller radii, suggesting that in the absence of any balancing heat
source much of the gas in the central regions is cooling out of the
hot intracluster medium. In order to maintain the pressure required to
support the weight of the overlying gas, a slow, subsonic inflow known
as a cooling flow develops.

X-ray observations made before Chandra and XMM-Newton were broadly
consistent with the cooling flow picture (see Fabian 1994 for a
review), although several issues remained unresolved. The first issue
was the observed X-ray surface brightness profile, which was not as
peaked as expected from a homogeneous flow. Instead a multiphase gas
was assumed, dropping cold gas over a range of radii.  The second was
the fate of the cooled gas. At the rates of 100s to more than
$1000\Msunpyr$ found in some clusters, the central galaxies should be
very bright and blue if the cooled gas forms stars with a normal
intial-mass-function. In many cases they do have excess blue light
indicative of massive star formation (Johnstone, Fabian \& Nulsen
1987; Allen 1995; Cardiel et al 1998; Crawford et al 1999; Bayer-Kim
et al 2002), but at rates which are a factor of 10 to 100 times lower
than the X-ray deduced mass cooling rate. It has also been argued
(e.g. O'Dea et al 1994) that there is no significant sink in terms of
cold gas clouds. A third issue involved the shape of the soft X-ray
spectrum, which was inconsistent with a simple cooling flow.
Absorption intrinsic to the flow was found to be a possible
explanation (Allen \& Fabian 1997; Allen et al 2001a).  A final issue
was whether the neglect of heating was justified. The effect of
gravitational heating as the gas flows was taken into account, but the
effects of any central radio source, which pumps energy into the
surrounding gas via jets, together with disturbances due to
subclusters plunging into the core every few Gyr were not included due
to a lack of quantitative information. Heat flow due to thermal
conduction was also assumed negligible.

The situation with cluster cooling flows has been clarified over the
past two years, particularly by the high spatial resolution imaging of
Chandra and the high spectral resolution of the XMM-Newton Reflection
Grating Spectrometer (RGS). Chandra images show much detail in the
cores of clusters, with bubbles from radio sources (e.g. McNamara et
al 2000; Fabian et al 2000; Blanton et al 2001) and cold fronts
(Markevitch et al 2000; Vikhlinin et al 2001) seen. RGS spectra
(Peterson et al 2001; Tamura et al 2001; Kaastra et al 2001) confirm
the presence of a range of temperatures in cooling flow clusters but
fail to show evidence of gas cooling below 1--2~keV. Simply put
the data are consistent with gas cooling at a high rate to about one
third of the gas temperature beyond 100~kpc but then vanishing.

At about the same time, the evidence for both warm (Jaffe \& Bremer
1997; Donahue et al 2000; Edge et al 2001; Wilman et al 2002) and cold
(Edge 2001; Salom\'e \& Combes 2002) molecular gas (H$_2$ and CO) at
the centres of cooling flows clusters has become widespread. In some
extreme cases there may be over $10^{11}\Msun$ of cold gas (Edge
2001).The presence of dust in these regions is also widespread, from
evidence of the Balmer decrement in the optical/UV nebulosities
commonly seen (e.g. Heckman et al 1989; Crawford et al 1999),
dustlanes, and submm and IR detections (Edge et al 1999; Allen et al
2001a; Irwin, Stil \& Bridges 2001). It is therefore possible that some
more star formation, and in particular cold gas clouds, may be found
in and around central cluster galaxies. There has also been the
detection of OVI emission from A2597 with FUSE at a level consistent
with a moderate cooling flow (Oegerle et al 2001). Lastly, recent
numerical simulations of evolving cluster which include radiative
cooling of the gas predict cooling flows (e.g. Pearce et al 2000).

\begin{figure}
\includegraphics[width=\columnwidth]{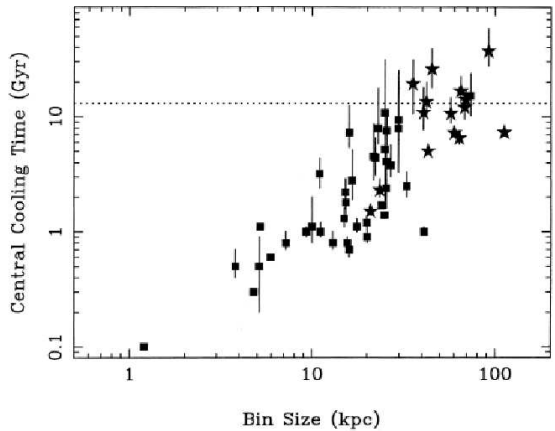}
\caption{Radiative cooling time inferred from ROSAT images of the
brightest 50 clusters in the Sky (Peres et al 1997). The bin size
depends on the quality of the data and is smallest for nearby peaked
clusters. Note that most clusters have central cooling times less than
a Hubble time (dashed line).  }
\end{figure}

\begin{figure}
\includegraphics[angle=270,width=\columnwidth]{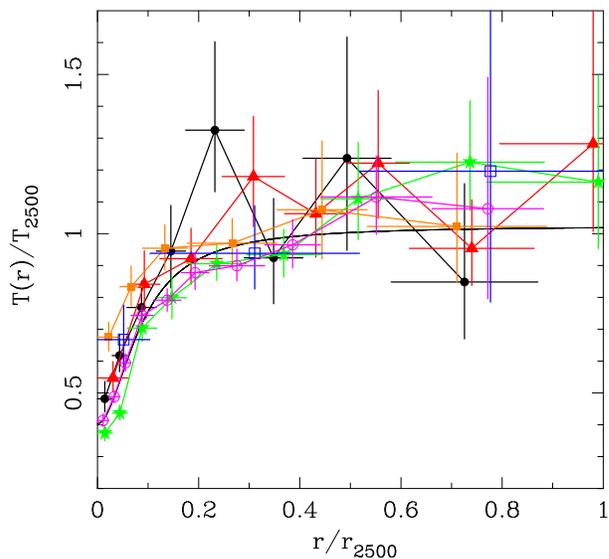}
\caption{Chandra temperature profiles for six massive clusters:
PKS0745-191, A2390, A1835, MS2137-2353, RXJ1347-1145 and 3C295, from
Allen et al (2001b). Note the temperature drop within $\sim0.2
r_{2500}\sim 120\kpc$. }
\end{figure}

Some form of heating may balance radiative cooling but the source of
heating remains unsolved, although several good candidates have been
identified (e.g. Tucker \& Rosner 1983; Binney \& Tabor 1995). Heating
from radio sources and infalling subclusters must occur at some level,
but whether they can balance radiative cooling over the required
spatial scales to better than a factor of a few is not yet clear.
Thermal conduction is another candidate. Cooling probably does account
for the observed star formation and cold gas clouds but this may be
confined to just the inner 10~kpc or so. Whether the mass cooling
rates are reduced from the earlier X-ray deduced rates by a factor of
a few, ten or a hundred is unclear.

Some form of feedback is probably required to prevent all of the gas
from being heated up. If feedback does occur we have a good chance to
observe how it works, since the region is spatially resolved and
optically thin. The process is of wide importance, since feedback
modulated cooling is a key ingredient in the baryonic part of galaxy
formation.

This is an updated and extended version of Fabian (2002d). 

\section{Chandra results}

Chandra images show structure in cluster cores. The X-ray emission is
steeply peaked into the centres of many clusters but there are holes
and fronts in the peak. Markevitch et al (2000) found sharply defined
cold fronts on A2142, across which the pressure is continuous yet the
temperature changes by a factor of about 2. Ettori \& Fabian (2000)
note that thermal conduction must be heavily suppressed at the fronts
in order that such sharp features can last long enough to be common.
The fronts probably indicate that the gas of subclusters does not
readily mix with the existing intracluster medium, presumably because
they are separate magnetic structures. Also the cores of infalling
subclusters need not be strongly shocked in decelerating into the core
(see Fabian \& Daines 1992).

\begin{figure}
\includegraphics[width=\columnwidth]{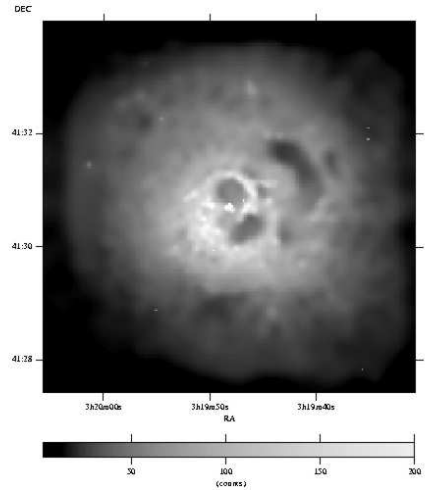}
\includegraphics[width=\columnwidth]{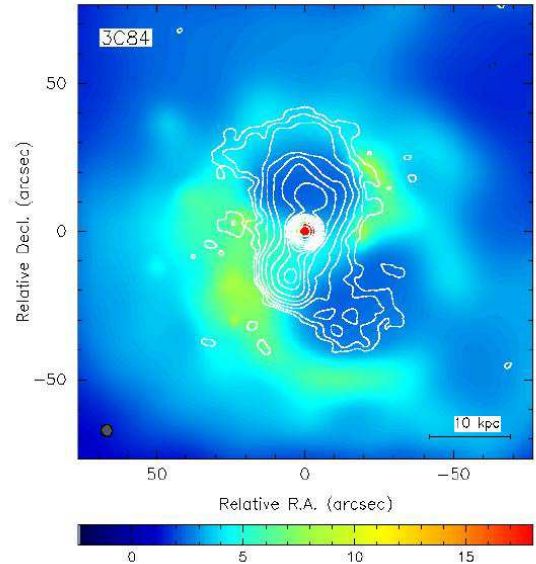}
\caption{(Top) Adaptively-smoothed 0.5--7~keV ACIS-S X-ray image of
the Perseus cluster; (Bottom) Radio image (1.4~GHz restored with a 5
arcsec beam, produced by G. Taylor; see Fabian et al 2000a) overlaid
on an adaptively smoothed 0.5--7~keV X-ray map.}
\end{figure}

\begin{figure}
\includegraphics[width=\columnwidth]{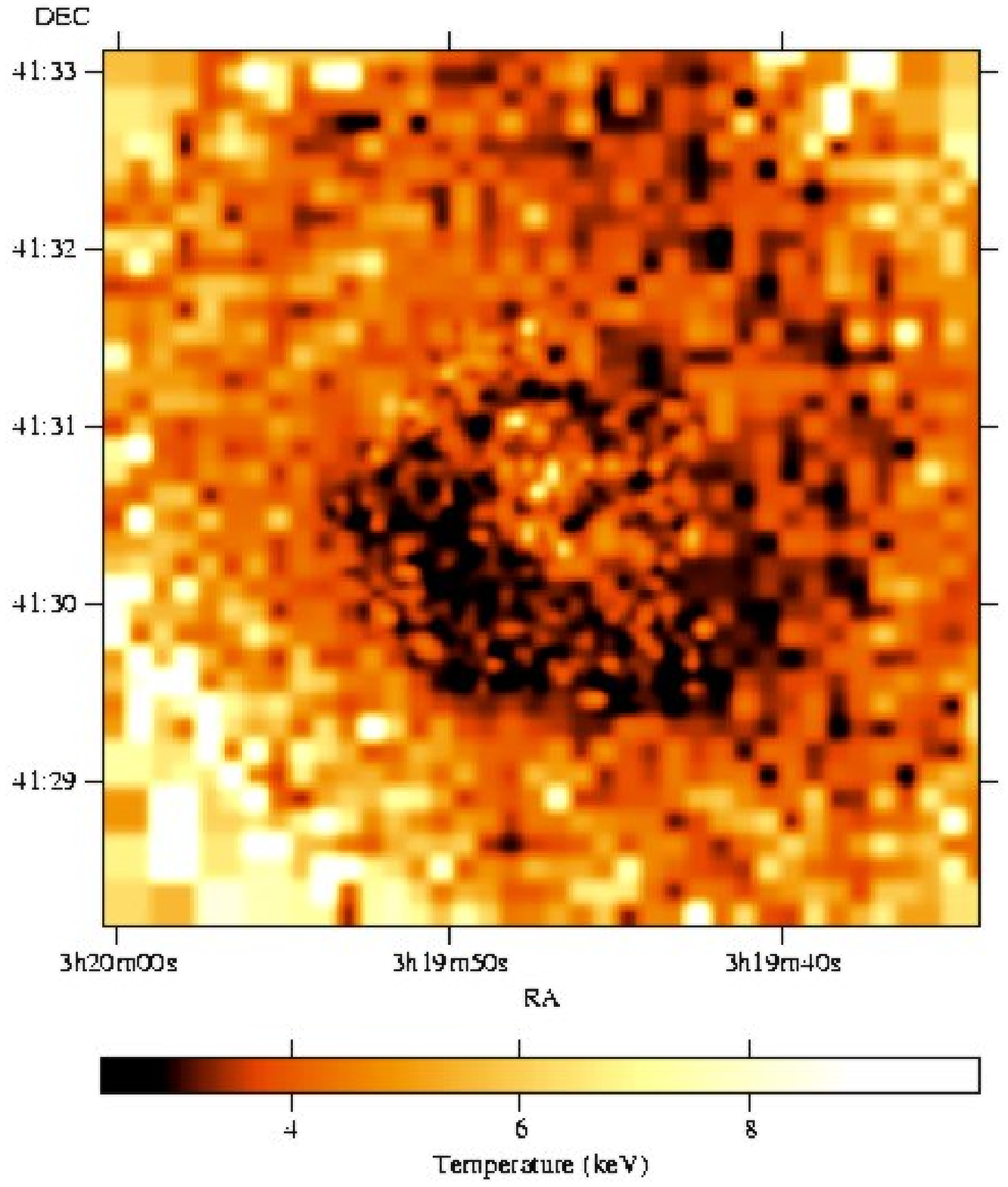}
\includegraphics[width=\columnwidth]{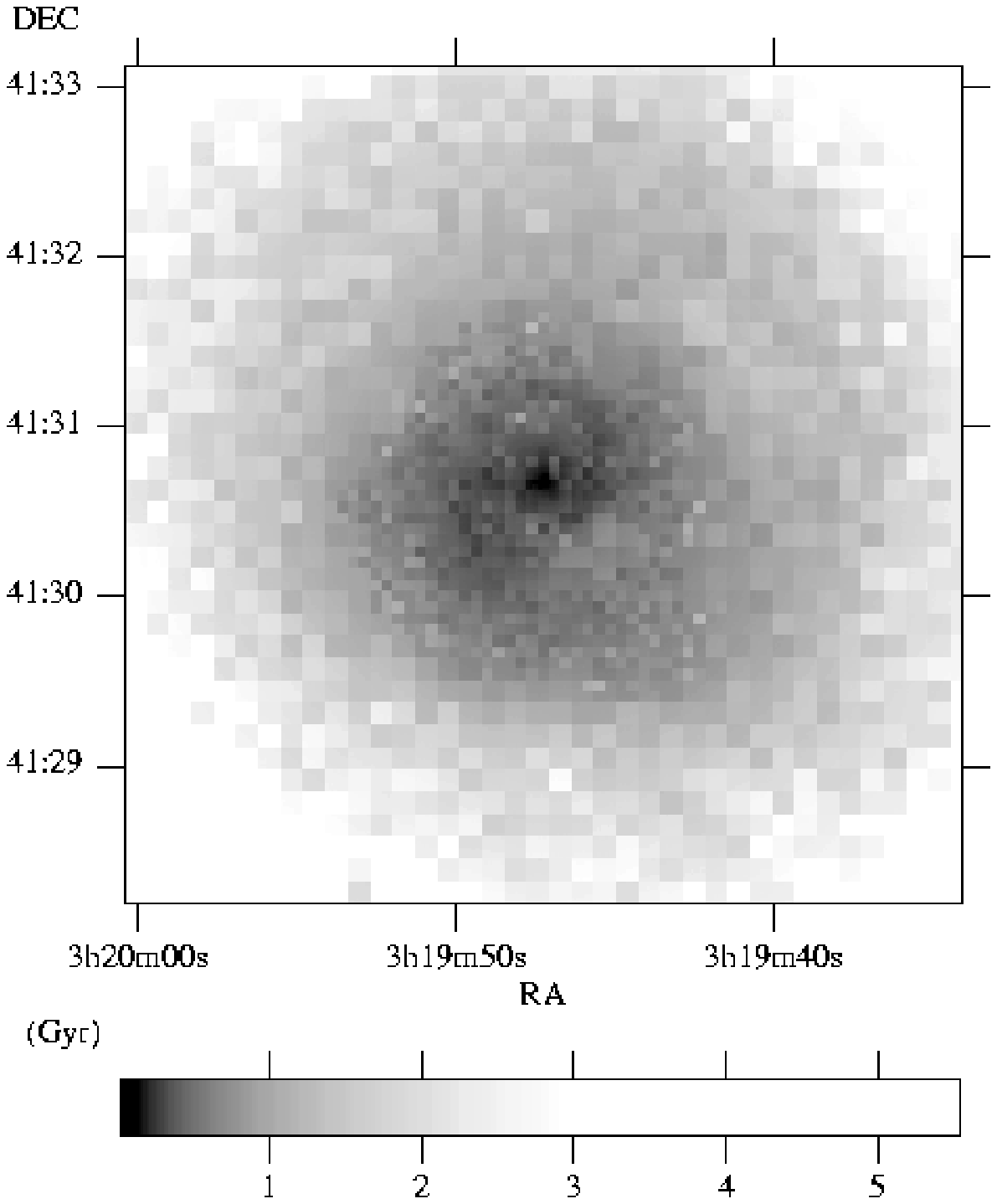}
\caption{(Top) Temperature and (Bottom) radiative cooling time maps of
the Perseus cluster (Fabian et al 2000a). Note that the coolest gas
($T\sim 2.5\keV$) with the shortest cooling time ($\sim 0.3$~Gyr) lies
in the rim around the N lobe, and the swirl to the temperature map.
Single-phase gas has been assumed for the analysis. }
\end{figure}

\subsection{The Perseus, Centaurus and Virgo clusters}

Holes in the X-ray surface brightness are seen to coincide with some
radio lobes. Good examples are in the Perseus cluster (Figs.~3 and 4),
and were first seen with ROSAT (B\"ohringer et al 1993). Chandra shows
that they have bright rims of X-ray {\it cool} gas (Fabian et al
2000a; Schmidt et al 2002). This is contrary to the work of Heinz et
al (2000) who predicted that the rims would signify shocks. Other
holes coincident with radio lobes are found in Hydra A (McNamara et al
2000; David et al 2001; Nulsen et al 2002) and many other clusters
(e.g. A2052, Blanton et al 2001; A2597, McNamara et al 2002). The
puzzling aspect if radio sources are heating the cooling gas is that
in all cases reported the {\it coolest} gas seen is that closest to
the radio lobes. Of course there is much energy going into the lobes,
but the energy from the $PdV$ work expended in forming the holes can
propagate away as sound waves, and the relativistic energy stored in
the bubbles (Fabian et al 2002a) can be lifted away and out of the
immediate core by buoyancy.

\begin{figure}
\includegraphics[width=0.9\columnwidth]{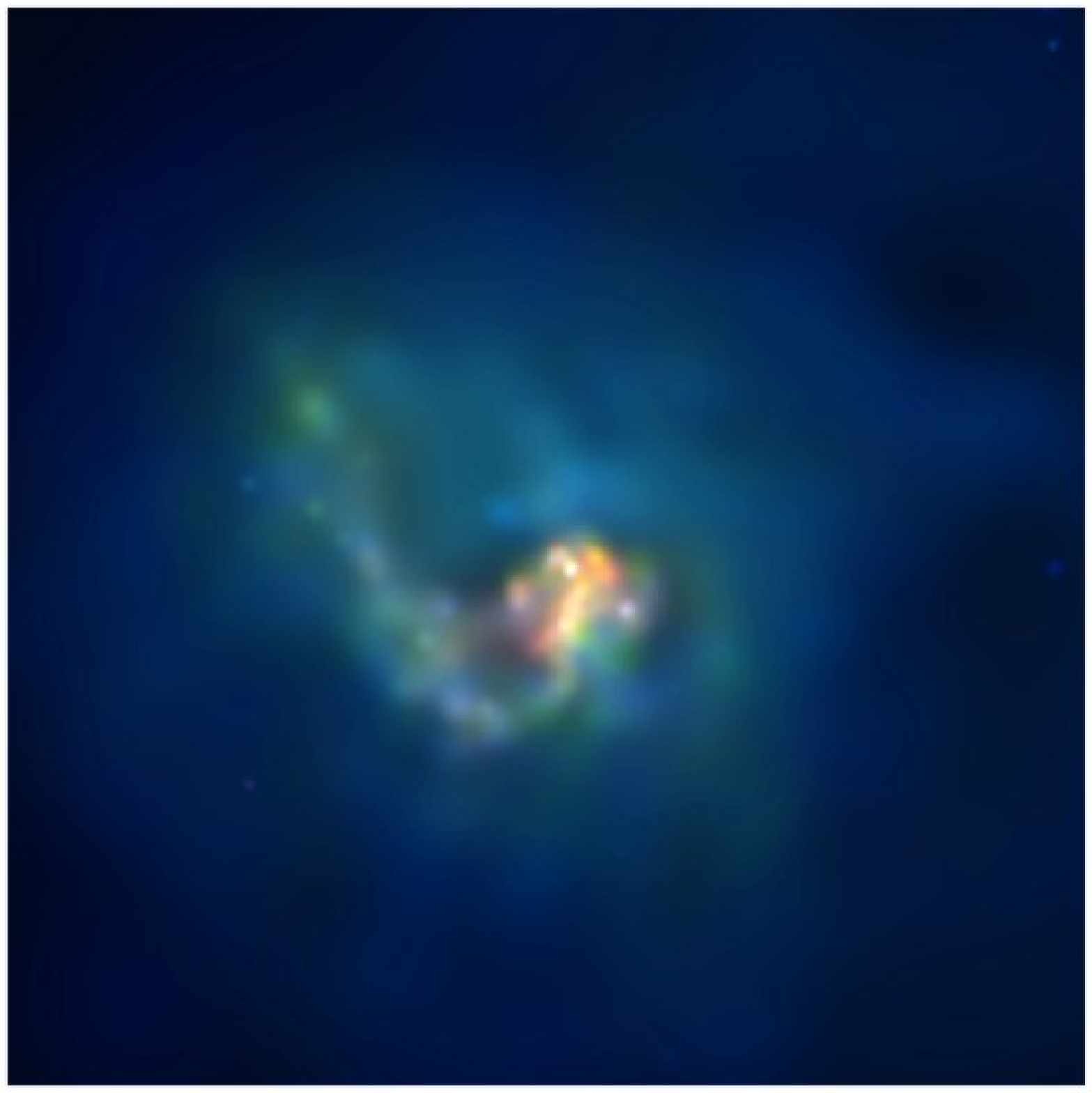}
\includegraphics[width=\columnwidth]{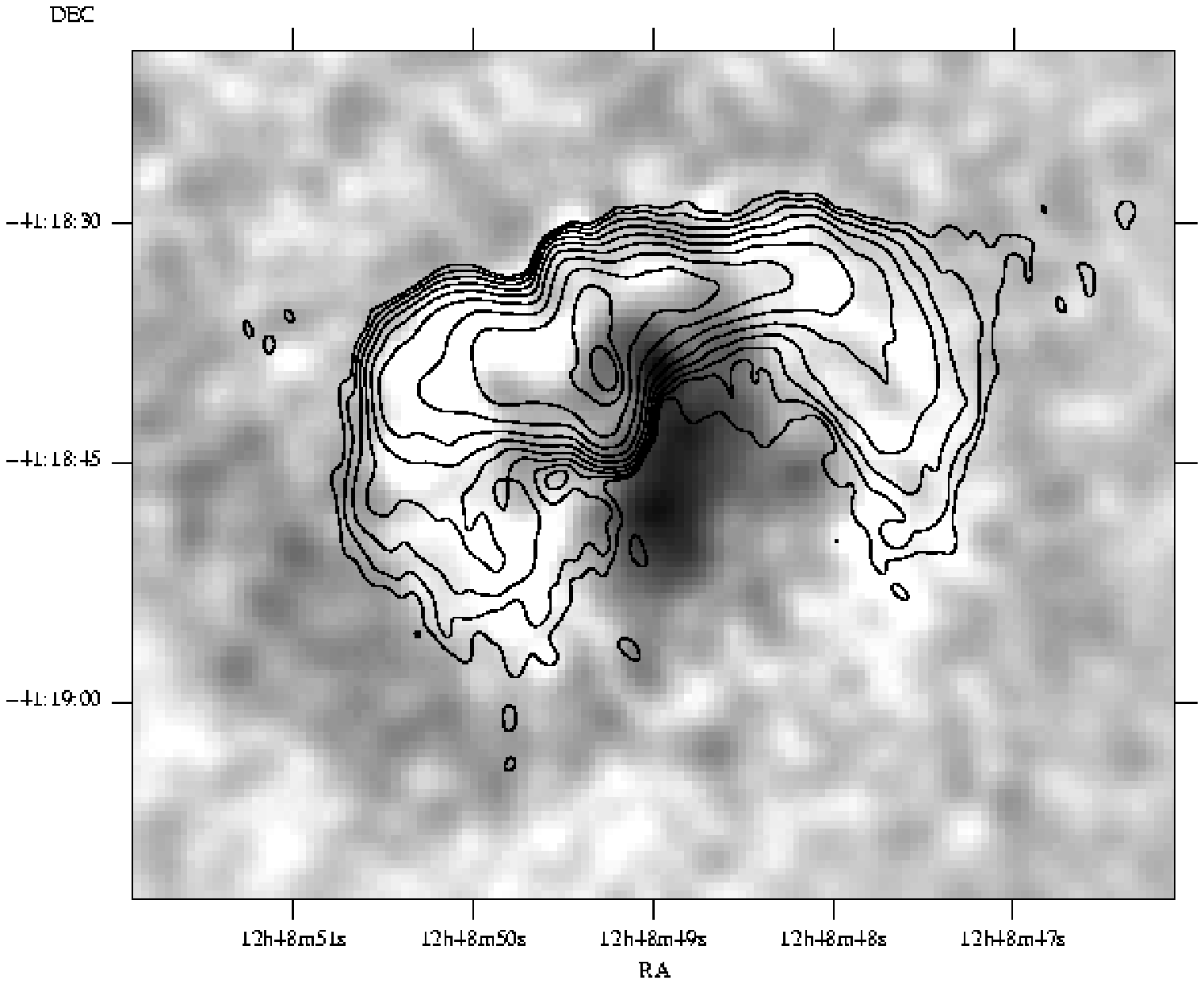}
\caption{(Top) Colour image of the Centaurus cluster, with red
indicating the coolest gas. (Bottom) radio contours on an unsharp mask
of the X-ray image, showing that the radio source has displaced the
X-ray gas.  }
\end{figure}

\begin{figure}
\includegraphics[width=\columnwidth]{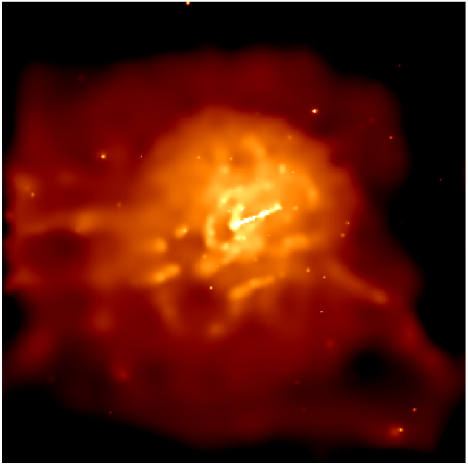}
\caption{40~ks Chandra image of the core of the Virgo cluster (Young,
Wilson \& Mundell 2002). The famous M87 jet is seen pointing at 2
o-clock in the centre. }
\end{figure}

The Centaurus cluster (Fig.~5; Sanders \& Fabian 2002) shows a swirly
structure of X-ray cooler gas. The radio source sits around the cooler
central blob, displacing the X-ray emission. Strong Faraday Rotation
is seen in the radio source (Taylor et al 2002), indicating
significant magnetic fields in the surrounding gas. Field
amplification could result from the inflow associated with a cooling
flow. 

The Virgo cluster (Fig.~6; Young, Wilson \& Mundell 2002) shows ridges
to the N of the nucleus coincident with optical filaments and some
bubble-like features. Cool structures extend along the outer radio
lobes to the E and SW (see also ROSAT and XMM-Newton; B\"ohringer et
al 1995; 2001). The Eastern one is likened to a mushroom cloud
following an explosion by Churazov et al (2001).

\subsection{A1795 and A2199}

\begin{figure}
\includegraphics[width=\columnwidth]{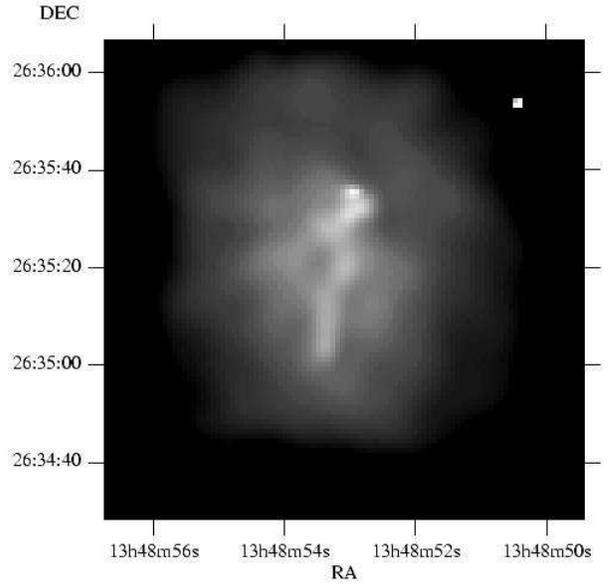}
\caption{Adaptively-smoothed X-ray image of the centre of A1795
(Fabian et al 2000b).}
\end{figure}

\begin{figure}
\includegraphics[width=\columnwidth]{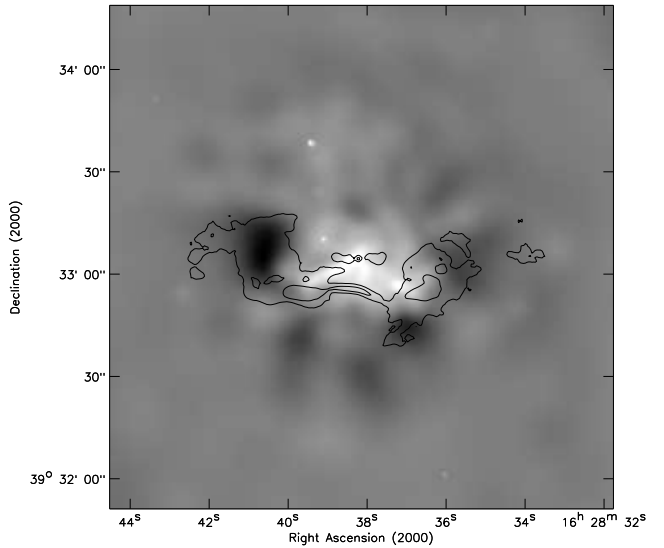}
\caption{X-ray image of A2199 with overlaid radio contours 
(Johnstone et al 2002). }
\end{figure}

\begin{figure}
\includegraphics[width=\columnwidth]{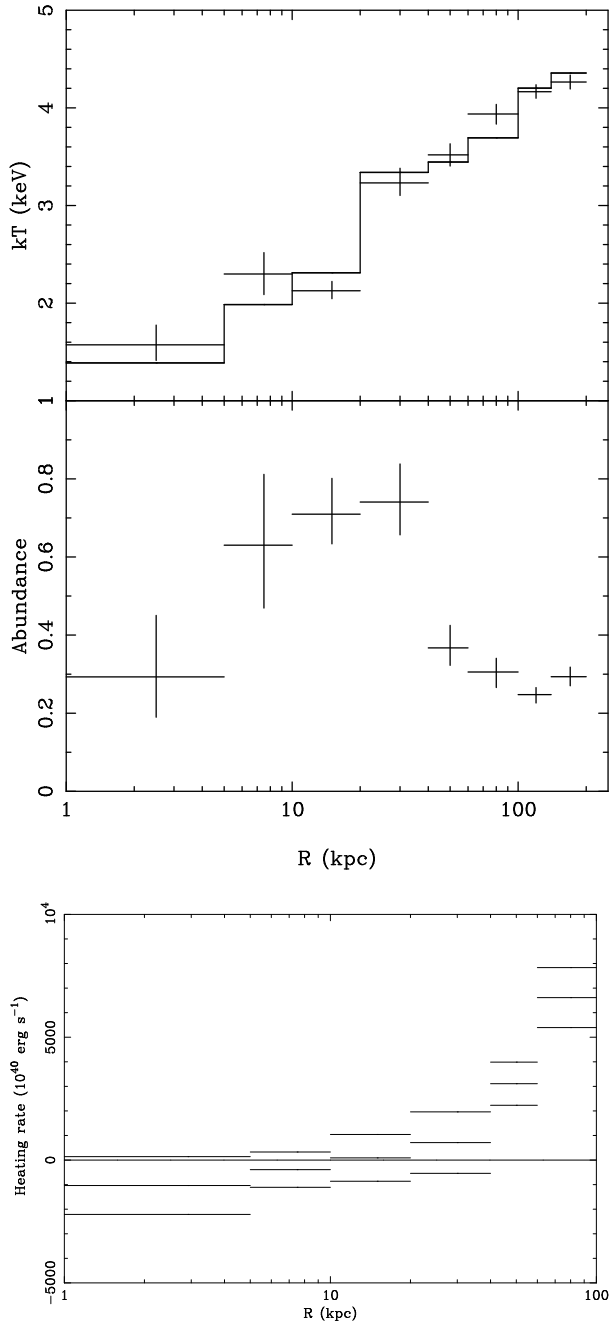}
\includegraphics[angle=270,width=\columnwidth]{afabianfig13.ps}
\caption{(Top) Deprojected temperature and (Middle) Abundance profiles
for A2199. (Bottom) Heating rate required to maintain mass flow rates
of 0 (top set of lines), 50 (middle set) and 100 (lower set)
$\Msunpyr$ in A2199.  }
\end{figure}

The Chandra image of A1795 (Fabian et al 2000b; Ettori et al 2002)
shows an 80~kpc long soft X-ray feature coincident with an H$\alpha$
filament found by Cowie et al (1982). It is plausibly a cooling wake
trailing behind the central galaxy, which is at the head of the
filament. The galaxy is moving around in the core of the cluster at a
few hundred $\kmps$. There is no evidence from the temperatures of the
gas that this motion has heated the gas significantly. The central
galaxy in A2199 may also be oscillating, as deduced from the unusual
morphology of its radio source (Burns et al 1983) and X-ray emission.
Again, the coolest gas appears to be close to both the radio source
and the central galaxy (Johnstone et al 2002).

The (deprojected) temperature and metal abundance profiles in A2199
are shown in Fig.~9 (from Johnstone et al 2002). As is now seen to be
typical for such objects, the temperature drops by about a factor of
three. The radiative cooling time of the gas is around $10^{10}\yr$ at
100~kpc, $7\times 10^8\yr$ at 10~kpc and $\sim 10^8\yr$ within 2 kpc.
The heating rate required to maintain the gas in thermal equilibrium
is also shown in Fig.~9. It is clear from this plot that the heating
must be distributed and cannot just occur at the very centre. What is
required is roughly equal amounts of power per kpc around the centre.

The emission measure distribution of the gas in the core of A2199 is
also compared with that for a cooling flow in Johnstone et al (2002).
This distribution increases with radius, whereas that expected from a
single-phase cooling flow is weakly decreasing. So although the
temperature profiles resemble those of cooling flows, the emission
measure distribution (and thus the gas density distribution) does not.

\section{XMM-Newton results}

\begin{figure}
\includegraphics[width=1.1\columnwidth]{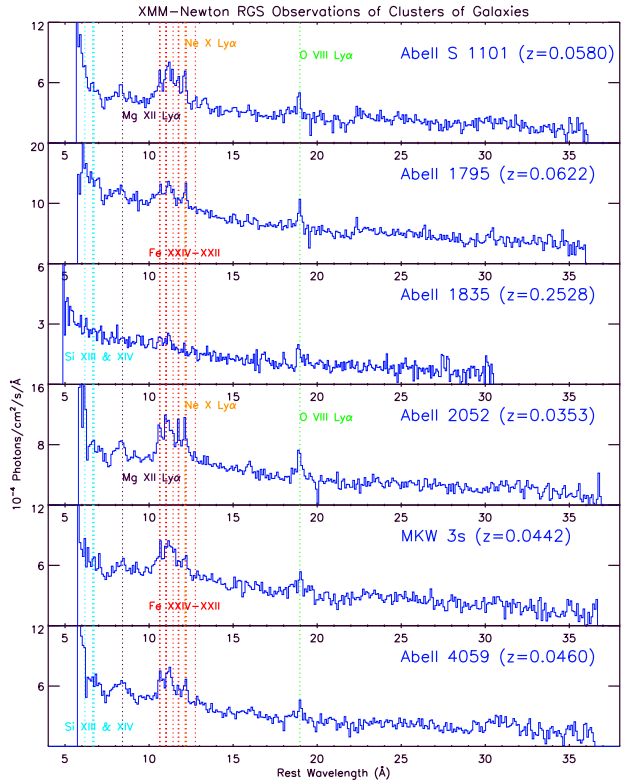}
\caption{RGS spectra of 6 cooling flow clusters, kindly provided by J.
Peterson and J. Kaastra. Emission lines between 10-13~A indicate the
presence of cooler gas in these clusters (at 1--3~keV) but the lack of
lines between 13 and 18~A shows that gas is not radiatively cooling
below 1--2~keV at high rates in any simple unobscured manner. }
\end{figure}

The most striking results have come from the RGS data which show
little evidence for gas cooling below 1--2~keV (Fig.~10; Peterson et
al 2001, 2002; Tamura et al 2001; Kaastra et al 2001). These grating
spectra of the inner 30 arcsec of cluster cores are at higher spectral
resolution than previous observations and clearly show the Fe-L lines
of Fe XXII-XXIV at 10--13A from gas at about one third the outer
cluster temperature. Emission lines from FeXX and XVII at 13--18A
should however be bright and easily seen if a continuous cooling flows
is taking place, but they are absent. EPIC CCD spectra (e.g.
B\"ohringer et al 2001; Molendi \& Pizzolata 2001) confirm this
result. For a range of objects the upper limits on the mass cooling rate
are about one fifth of that previously proposed.

In summary the data for a typical relaxed cluster show that the
radiative cooling time of the gas is about $10^{10}\yr$ at 100-200~kpc
from the centre, decreasing monotonically by one to two orders of
magnitude inward to the centre. The gas temperature decreases to the
centre by at least a factor of three. The spectrum of the whole region
is consistent with gas cooling steadily from the outer temperature to
about one third of that value at a rate which may be high ($\dot M\sim
100-1000\Msunpyr$ in massive clusters), but the limit on gas cooling
below one third of the outer temperature is only 20 per cent of that
level of $\dot M$. The surface brightness is not as peaked as expected
from a single phase cooling flow (i.e. a single temperature at each
radius) but would require a multiphase flow (i.e. a range of
temperatures at each radius). There is little evidence for any extreme
multiphase gas with temperatures ranging over an order of magnitude
(see Molendi \& Pizzolato 2001 and Johnstone et al 2002 for limits on
multiphase gas) but it would be difficult to rule out a factor of two
range variation.

It seems likely that the temperature drop is due to cooling (although
gas from incoming cool subclusters may contribute). This raises the
problem of why the gas does not appear to cool below about one third
of the outer temperature.

\section{Heating and Cooling}

Various explanations have been discussed (e.g. Peterson et al 2001;
Fabian et al 2001; 2002b; B\"ohringer et al 2002). They can be split
according to whether heating plays a major role or not.

\subsection{Solutions without heating}

The gas may be cooling and yet appear to vanish when it reaches say
2~keV. Clouds of cold gas may for example photoelectrically absorb the
soft X-rays. So far no exhaustive testing of this possibility has been
carried out but it is unlikely to work unless the absorbing gas is
intimately related to the cooling gas blobs.

The gas may have become dense enough to separate from the flow and mix
in with surrounding hotter gas (Norman \& Meiksin 1996).  In this
category are also some models involving a central radio source which
is merely required to drag the coolest gas from the very centre out to
larger radii (David et al 2001; Nulsen et al 2002). Such rearrangement
of the gas does not of course change the long term cooling rate of the
whole region. Taking the cooler gas rapidly to larger radii may
adiabatically cool it, possibly reducing the observed emission from
the 0.2--1~keV temperature range, but the gas would still cool (for
bremsstrahlung the cooling time goes as pressure$^{-0.4}$ and is
roughly constant with pressure for line dominated cooling; the
pressure changes by about a factor of 10 over the inner 100kpc). Only
if the gas can be dragged out well beyond the inner 100~kpc would the
total cooling flow rate to below $10^6\K$ be significantly reduced.

Alternatively it may mix in with colder gas (for example that
associated with the optical filaments or the molecular gas, which
would explain why the filaments are so bright; Fabian et al 2002b).
The problem is then best seen as one of missing soft X-ray luminosity
(i.e. that of the X-ray coolest gas). The missing luminosity may
emerge in the FUV or infrared bands.

Another possibility is that the metals in the gas are not uniformly
mixed in, but have a bimodal distribution (Fabian et al 2001; Morris
\& Fabian 2002). Gas in which ten per cent has a metallicity of 3
times solar and 90 per cent has zero metallicity has the same spectrum
as gas at 0.3 solar if cooling is unimportant. When it does cool
however, the metal emission lines cool only 10 per cent of the gas and
so are much reduced as compared with the situation if they were
responsible for cooling all the gas. This model has not yet been
thorughly tested (assuming that the inhomogeneities occur on scales
much less than a kpc). It could  account for the puzzling central
metallicity decrease seen in A2199 (Fig.~9, Johnstone et al 2002) and
in the Centaurus cluster (Sanders \& Fabian 2002), since the
innermost gas will have been depleted of metal rich gas. 

Solutions with little heating, at least within the inner tens of kpc,
remain of interest as they seem best able to explain the FUSE
detection of OVI (Oegerele et al 2001) in A2597.

\subsection{Solutions with heating}

Two main heating solutions have been proposed, making use of two large
sources of energy; a central black hole and the outer cluster gas. The
first uses a central radio source, which is a common constituent of
such clusters and the other invokes conduction of heat from the outer
parts of a cluster.

First let us look at the level of heating required. It cannot just be
some low level of heat which stops the gas at 1--2~keV, since that
would cause an accumulation at that temperature, contrary to
observation. It has to halt the cooling over the full range of
temperatures, and thus radii. How this can happen is a puzzle. If the
radio source is responsible, then it may be intermittent. Maybe we do
not see the heating phase, which is short lived. The power required to
stem the flow during the heating phase then goes up to high values.
There may not be any problem in the Virgo cluster around M87
(e.g.Churazov et al 2002) where the
energy requirements are relatively small, but in a massive cluster
like A1835 (Peterson et al 2001; Schmidt et al 2001) the necessary
power may exceed $10^{46}\ergps$.

\begin{figure}
\includegraphics[angle=270,width=\columnwidth]{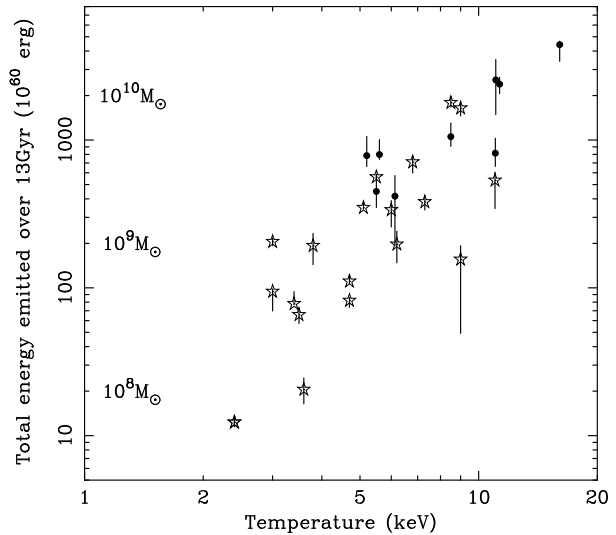}
\caption{Energy requirement in order to balance the radiative cooling
of cluster cores (Fabian et al 2002). The energy radiated within the
region where the cooling time is $10^{10}\yr$ is shown for about 25
clusters. The total black hole mass accumulated assuming an efficiency
of 10 per cent is indicated. }
\end{figure}

\begin{figure}
\includegraphics[angle=270,width=\columnwidth]{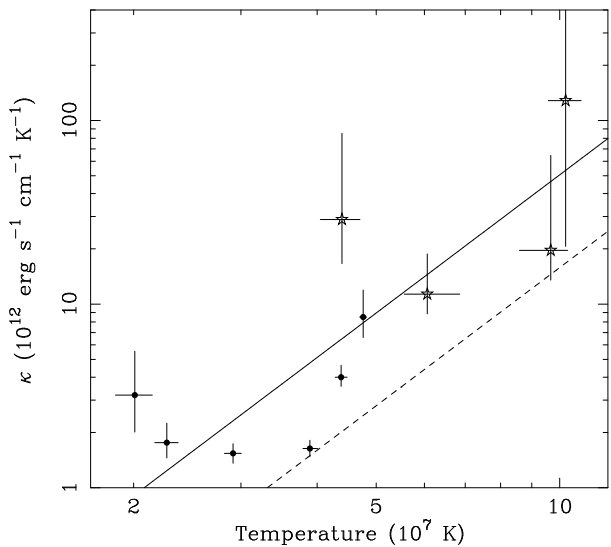}
\caption{The effective conductivity required to balance cooling in
A2199 (solid circles) and A1835 (stars; Voigt et al 2002). The solid
line represents conduction at the Spitzer rate and the dashed one is
one third of that rate. }
\end{figure}
\begin{figure}
\includegraphics[angle=270,width=\columnwidth]{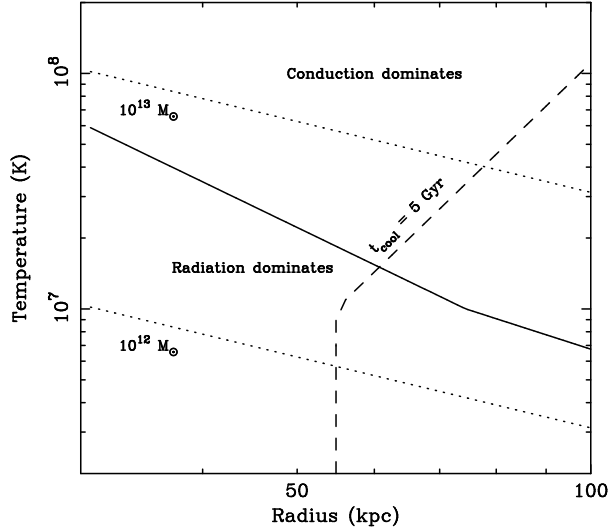}
\caption{Conduction-dominated regime, where conduction can balance
radiative cooling, for gas in an isothermal dark matter potential
well (Fabian et al 2002). The gas fraction is 10 per cent.  }
\end{figure}

The energy requirement is shown in Fig.~11 (Fabian et al 2002c). For
massive high temperature clusters, energy balance requires that most
of the accretion energy of a massive central black hole go into
heating the surrounding gas. This probably has to happen in a
radiationless manner around the black hole, with little of the
accretion energy appearing as radiation which couples very poorly with
the intracluster medium, and most of it as jet energy which is then
required to couple well. Radiation-inefficent flows such as ADAFs
(e.g. Narayan et al 1995) are not appropriate for such large implied
accretion rates (this is a problem for the feedback model for M87 of
Churazov et al 2002, which requires that at least 99 per cent of the
accretion energy goes into a jet). Only if the energy being tapped is
the spin energy of the black hole, could the low radiation, high jet
power, constraints be met.

Radio source models must do the heating in a manner that leaves the
surroundings of the inner radio bubbles cool. Various simulations have
shown that some heating does take place, and have presumed that it is
sufficient (Churazov et al 2000; Quilis et al 2001; Br\"uggen \&
Kaiser 2001; Reynolds et al 2002). There seems to be enough power in
the FRI jets to balance the cooling in modest cooling flows (as noted
by e.g. Pedlar et al 1990) if it can be effectively targetted. What
has yet to be shown is that the jet power can heat the intracluster
medium in a distributed manner leaving it looking like observed
clusters. If too much power is used then presumably the jets break
through as in Cyg A and dump most of their power at large radii beyond
the cooling region.

There is no doubt that the radio sources push the gas around and ought
to do some heating, but the level of heating has yet to be determined.
Note that the gas must not be stirred to the extent that the abundance
gradients are destroyed. Feedback models where some accretion is
followed by a heating phase seem most promising.

The other large heat source is the outer intracluster medium. The
cooling region within 100~kpc is only a few per cent of the total mass
of gas in a cluster and the outer gas is hot. Narayan \& Medvedev
(2001) have noted that the level of conductivity required to offset
cooling is similar to the Spitzer value expected in a plasma. Indeed,
they have shown that in a turbulent magnetized plasma a value of
about one quarter of the Spitzer one could be appropriate.

Voigt et al (2002) have used deprojected temperature and emission
measure profiles of clusters to determine the effective conductivity
required. Interestingly for A2199 and A1825 it is less than the
Spitzer value over a significant temperature (and thus radius) range
(Fig.~12). Further work by Zakamska \&  Narayan (2002) indicates that
there are clusters for which Spitzer conductivity may not be
sufficient. 

Conduction is suppressed in cold fronts and may be suppressed
throughout most cluster cores. Different parts of the gas may have for
example different magnetic structures. Conduction may be time variable
and intermittent. Perhaps an intital cooling flow leads to a radial
magnetic field structure which enables efficient conduction to
operate. Conduction is an interesting, but unproven, component.

\section{Galaxy formation}

Much of the gas in the formation of massive galaxies behaves as in a
cooling flow (White \& Frenk 1991). Some form of feedback is required
to prevent everything from cooling at the earliest times, and cooling
must be switched off in the most massive objects (Kauffmann et al
1999). If we cannot understand what is going on in cluster cores, then
why should we believe models for galaxy formation which rely on a
similar process?

The feedback in galaxies is assumed to be due to supernovae, but it is
plausible that a central black hole is also important, especially
given the correlations between black hole and galaxy mass (Gebhardt et
al 2000; Ferrarese \& Merritt 2000). Only a few per cent of the
accretion energy of a central black hole is sufficient to unbind the
host galaxy. It is then puzzling why cooling should proceed in
galaxies (giving their stellar mass) but fail in clusters. 

The conduction solution to cooling flows gives one answer to this,
since conduction works best at high temperatures, whereas cooling
works best at low temperatures. The regimes where conduction and
radiative cooling dominate in gas trapped in an isothermal dark matter
well are shown in Fig.~13 (from Fabian et al 2002c). This shows that
if conduction does proceed unsuppressed in galaxy formation then it
moulds the upper mass limit for galaxies.

\section{Summary}

The central 100 kpc radius region in most clusters has a radiative
cooling time shorter than 5 Gyr and many have cooling times which drop
to the centre to a value of only $10^8\yr$ or so. The gas temperature
drops by a factor of 3 or more over this radius range. It is
plausible that the temperature drop and short radiative cooling times
are related and that the low temperatures are caused by radiative
cooling.

It is then a puzzle as to why gas which has lost two thirds of its
thermal energy, and for which the radiative cooling time is very
short, is not seen to cool further. 

There are two obvious solutions, both of which have some difficulties.
The first solution is that the gas does cool but either the soft X-ray
emission is absorbed or the cooling is non-radiative and due, say, to
mixing. The problem of the fate of cooled gas then remains, although
significant masses of cold gas have been found in some objects. The
second solution is that some heating balances cooling. The problem
here is that the heat has to balance cooling over a wide range of
radii and a wide range of timescales. Observations of radio lobes
which are a likely source of heat indicate that they coincide with the
coolest gas in cluster cores. Heating by conduction is promising for
some clusters, although it requires that conduction proceed at close
to the Spitzer rate.

The data are consistent with modest levels of cooling ($\dot M$ of
tens rather than hundreds $\Msunpyr$) continuing in the inner tens kpc
in many clusters but with some heating, perhaps intermittent,
balancing most of the radiative cooling at large radii.

\section{Acknowledgements}
I thank the Organisers for a great meeting in such a lovely
place, and my colleagues, Roderick Johnstone, Jeremy Saunders, Robert
Schmidt, Glenn Morris, Lisa Voigt, Steve Allen and Carolin Crawford,
for help and discussions. The Royal Society is thanked for support.

\end{document}